\documentclass[
UTF8,
reprint,
nofootinbib,
nobibnotes,
bibnotes,
amsmath,amssymb,
aps
]{revtex4-2}

\usepackage{pdfpages}

\makeatletter
\AtBeginDocument{\let\LS@rot\@undefined}
\makeatother

\includepdfset{turn=false}
\usepackage[normalem]{ulem}
\usepackage{graphicx}
\usepackage{dcolumn}
\usepackage{bm}
\usepackage{hyperref}
\usepackage{subfigure}
\usepackage{xcolor}

\UseRawInputEncoding
\usepackage{amsmath}
\usepackage{amsfonts}
\usepackage{bbm}
\def\bv{\boldsymbol}
\def\im{\mathbbm{i}}

\def\ss{\scriptscriptstyle}
\begin{document}

	\preprint{APS/123-QED}

	\title{Hamiltonian with Energy Levels Corresponding to Riemann Zeros}

	\author{Xingpao Suo}
	\email{xpsuo@zju.edu.cn}
    \affiliation{College of Physics and Information Engineering, Zhaotong University, Zhaotong, Yunnan 657000, China }
	\affiliation{Institute for Astronomy, School of Physics, Zhejiang University, Hangzhou 310027, China             }
	
	\date{\today}

	\begin{abstract}
	   A Hamiltonian with eigenenergy \( E_n = \rho_n(1 - \rho_n) \) has been constructed, where \( \rho_n \) denotes the \( n \)-th non-trivial zero of the Riemann zeta function. To construct such a Hamiltonian, we generalize the Berry-Keating paradigm and encode number-theoretic information into the Hamiltonian using modular forms.Although our construction does not resolve the Hilbert-P\'olya conjecture (since the eigenstates corresponding to \( E_n \) are \emph{not} normalizable), it provides a novel physical perspective on the Riemann Hypothesis (RH). In particular, we propose a physical interpretation of RH, which could offer a potential pathway toward its proof.
	\end{abstract}

	\maketitle

	\textit{Introduction}-The Riemann zeta function, defined by the series \( \zeta(s) := \sum_{n \geq 1} n^{-s} \), lies at the core of analytic number theory. For instance, the growth order of many arithmetic functions (e.g., the prime number counting function \( \pi(x) \)) depends on the analytic properties of \( \zeta(s) \) \cite{Riemann1859, Titchmarsh1987}. Among the properties of \( \zeta(s) \), the distribution of its zeros is of utmost importance and interest. The famous Riemann Hypothesis (RH) asserts that \cite{Riemann1859}
	\begin{align}
		\left\{ 
		\begin{array}{c}
			\zeta(s) = 0  \\
			0<\Re s<1
		\end{array}
		\right.
		\Longrightarrow  \Re s = \frac{1}{2}\ ,  
	\end{align}
	where the condition \( 0 < \Re s < 1 \) excludes the so-called trivial zeros of \( \zeta(s) \), which occur at the negative even integers.

	Based on the RH, many related problems have been proposed. A physically motivated example is the Hilbert-P\'olya conjecture (HPC), which, in physical terms, asks whether there exists a self-adjoint Hamiltonian \footnote{For simplicity, in this paper the term \textit{Hamiltonian} implicitly refers to the quantum Hamiltonian. In formulas, we use a hat symbol ( \(\hat{}\) ) to emphasize its operator nature, thereby distinguishing it from the classical Hamiltonian which is a function. } whose energy levels satisfy  
    \(
    E_n = P(\rho_n) \quad \text{for all } n \in \mathbb{N}^+.
    \)  
    Here \(\rho_n\) denotes the \(n\)-th non-trivial zero of the Riemann zeta function with \(\Im \rho_n > 0\), and the function \(P\) must satisfy the condition  
    \(
    P(\rho_n) \in \mathbb{R} \iff \Re \rho_n = \frac{1}{2},
    \)
    for example, \( P(\rho_n) := -\im \left(\rho_n - \frac{1}{2}\right) \) or \( P(\rho_n) := (1-\rho_n)\rho_n \) \cite{MR337821}. 
    
    The HPC implies the RH; therefore, it is regarded as a potential pathway toward proving the RH and has attracted significant attention from many researchers in physics \cite{BK1999, German2007, German2008, German2011, Rev2011, Bender2017, LeClair2024}. For instance, Michael Berry and Jonathan Keating proposed the classical Hamiltonian \( H_\text{BK} := xp \) \cite{BK1999}, with the aim of quantizing it (under an appropriate scheme) to obtain a quantum Hamiltonian that satisfies the requirements of the HPC \cite{BK1999book}, where \( x \) and \( p \) denote the canonical coordinate and momentum, respectively. Although naive quantization of \( H_\text{BK} \) fails to achieve this objective \cite{German2011}, it has nonetheless inspired a large body of subsequent work, in which researchers have attempted to quantize or extend \( H_\text{BK} \) using various techniques \cite{German2007, German2008, German2011, Bender2017}. To date, however, these efforts have not succeeded in fully satisfying the requirements of the HPC.

	We now propose a Hamiltonian by extending the Berry-Keating framework from a different perspective. By encoding number-theoretic information into its structure, our Hamiltonian admits a novel set of energy levels \( E_n = \rho_n(1 - \rho_n) \). Although our model does not fully satisfy the HPC---since the corresponding eigenstates are not bound states---it nevertheless offers a fresh perspective on both the RH and the HPC.

	\textit{The model}-
	Motivated by the Berry-Keating paradigm, we introduce the following two-dimensional classical Hamiltonian:
    \begin{align}
         H = V p^2 \ ,  \label{equ:cham} 
    \end{align}
	where $p^2 = p_x^2 + p_y^2$ with $p_x$ and $p_y$ the momenta along the $x-$ and $y-$ axes, respectively, and $V(x, y) $ is a spatially dependent function, which we refer to as \textit{geometric potential}. 
	The similarity between our Hamiltonian and \( H_\text{BK} \) lies in the fact that both are products of coordinate-dependent and momentum-dependent quantities. However, in our case we use \( p^2 \) instead of \( p \) to avoid singularities in the system. Moreover, unlike the one-dimensional Berry--Keating Hamiltonian, the two-dimensional nature of our Hamiltonian allows for the existence of classical closed trajectories \cite{German2008}. This can be explicitly demonstrated by choosing \( V(x, y) = x^2 + y^2 \), for which the Hamiltonian's canonical equations admit circular orbit solutions. 
    
    Note that Hamiltonian \eqref{equ:cham} can be realized in specific physical systems, such as a heavy fermion with a position-dependent effective mass \cite{Kondo1964, Maja2019}, and a particle moving on a conformally flat surface with the metric \( g_{ij} = \frac{1}{V}\delta_{ij} \) \cite{DeWitt1957}.

	The quantization of Hamiltonian \eqref{equ:cham} faces an ordering ambiguity. Therefore, it is natural to employ a covariant quantization scheme, specifically the path-integral approach, which yields a formally Hermitian(symmetric) Hamiltonian \cite{DeWitt1957, Dekker1980, Assirati2017}:
    \begin{align}
        \hat{H} = - V^{\frac{1}{2}} \hat{\Delta} V^{\frac{1}{2}} \ ,  \label{equ:qham}
    \end{align}
    where \( \hat{\Delta} := \partial_x^2 + \partial_y^2 \) is the Laplace operator.

	Notably, our Hamiltonian exhibits fundamental differences from the conventional one, \( H_C := p^2 + V \), in two critical aspects, even though both models describe a particle moving in a \textit{potential} field.

    First, the behavior of particles differs significantly where the potential tends to infinity. Consider a potential that asymptotically diverges at infinity. For the system governed by \( H_C \), the particle is confined to a finite region because the momentum \( p \) will ultimately decrease to zero as the particle moves to the surface \( V(x, y) = E \), where \( E \) is the energy of the system. Quantum mechanically, this implies that the states governed by \( \hat{H}_C = -\Delta + V \) are always bound states.

    In stark contrast, for the system described by Hamiltonian \eqref{equ:cham}, the particle's momentum \( p \) never vanishes, even though it diminishes as \( V \) increases. This means that there always exist unbounded orbits. At the quantum level, we therefore expect the presence of scattering states, although their wave functions' modulus becomes very small near infinity.

    Second, our model exhibits singularities at the zeros of the geometric potential \( V \). Classically, the vanishing of \( V \) implies that the momentum \( p \) must diverge to infinity. Consequently, in the quantum regime, a physically admissible wave function should vanish at the zeros of \( V \) to satisfy regularity conditions.

	\begin{figure}
		\centering
		\includegraphics[width=1 \linewidth]{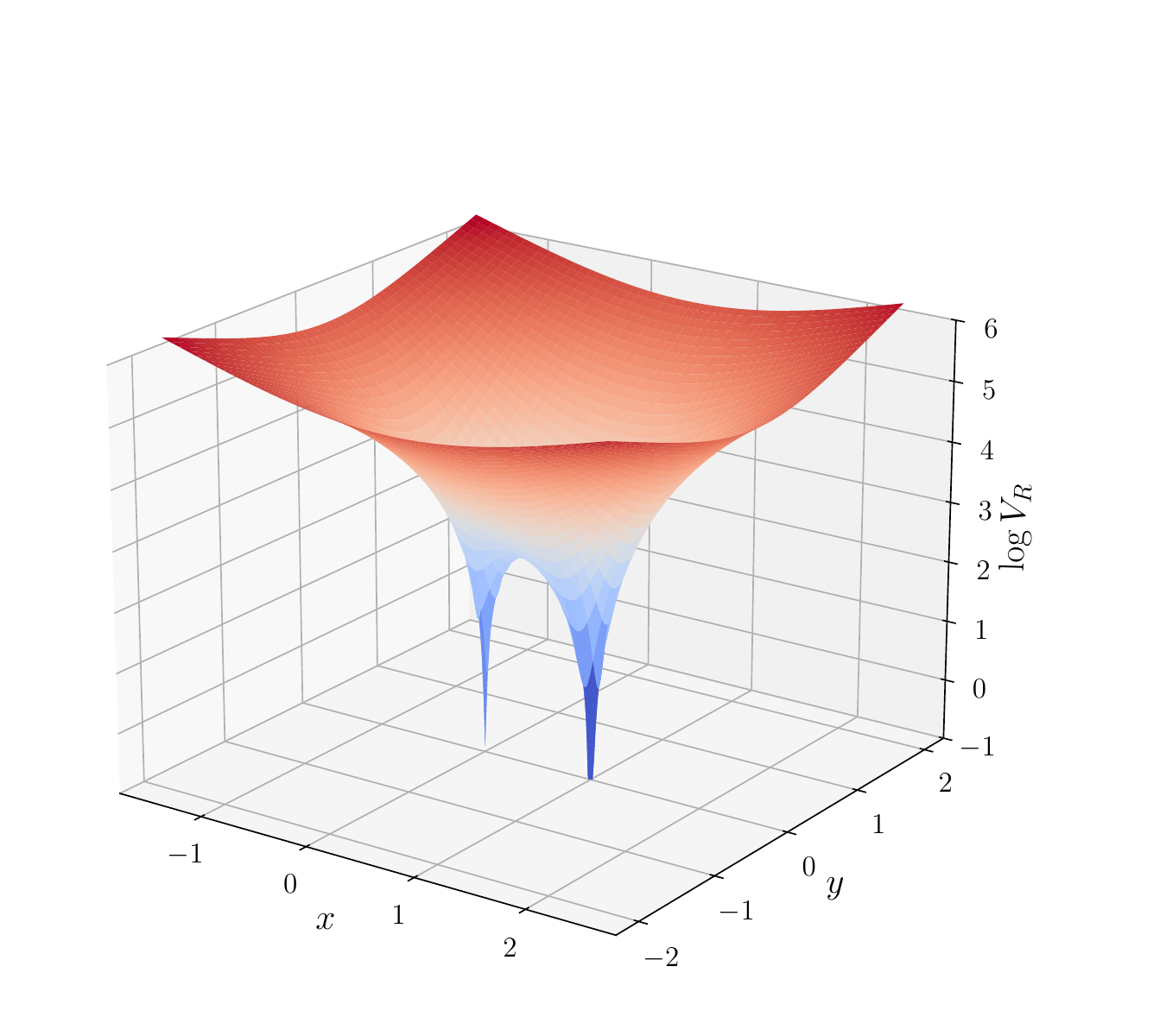}
		\caption{The double well structure of the geometric potential $V_{\ss R}(x, y)$. It can be clearly seen that there are two potential wells located at the origin and $(x, y) = (1, 0)$. Note here to demonstrate the structure of $V_{\ss R}$ we actually plot the logarithm of it.}
		\label{fig:Vxy}
	\end{figure}

	Our goal is now to construct a geometric potential $V$ that yields a set of eigenenergies corresponding to the Riemann zeros. Previous studies have provided several critical insights. For example, the statistical similarity between the distribution of zeta zeros and the eigenvalues of the Gaussian Unitary Ensemble (GUE) suggests that the desired geometric potential should correspond to a chaotic system \cite{MR337821, Odlyzko2001, Bogomolny1995}.  Moreover, to "encode" the information of the zeta function into the Hamiltonian, we need to consider a geometric potential that possesses a number-theoretic structure \cite{Bender2017}. Finally, since the zeta function satisfies the functional equation \( \xi(1 - s) = \xi(s) \) with \( \xi(s) := \pi^{-\frac{s}{2}} \Gamma\left(\frac{s}{2}\right) \zeta(s) \), the required geometric potential must reflect this symmetry \cite{Dobner2020, Maass1949}.

	For the above reasons, we consider the following geometric potential 
	\begin{align}
		V_{\ss R}(x, y) = \left( \frac{ \Im\tau(z)}{|\tau'(z) |}\right)^2  \label{equ:VP}
	\end{align}
	with $z := x + \im y \in \mathbb C $, $\im := \sqrt{-1}$, $\tau'(z) = d \tau(z)/dz $ and 
	\begin{align}
		\tau(z) := \im \frac
		{_2F_1\left(\frac{1}{6}, \frac{5}{6}, 1; \frac{1}{2} \left( 1 + \sqrt{\frac{z}{z-1}} \right) \right)}
		{_2F_1\left(\frac{1}{6}, \frac{5}{6}, 1; \frac{1}{2} \left( 1 - \sqrt{\frac{z}{z-1}} \right) \right) }\ 
		\label{equ:z2tau}, 
	\end{align}
	where $_2F_1(a, b, c; z)$ is the Gauss hypergeometric function. Note that first,  for any function $f(z)$, we restrict the argument range of $z$ to $(-\pi, \pi] $. Second, for convenience, we will use both complex and two-dimensional real notations. Specifically, we always assume \( z = x + \im y \) and write \( f(z) := f(x, y) \) for any function \( f(x, y) \) defined on \( \mathbb{R}^2 \). For example, we write \( V_{\ss R}(0) = V_{\ss R}(0, 0) \) and \( V_{\ss R}(1) = V_{\ss R}(1, 0) \).

	This unconventional geometric potential meets our requirements for the following reasons. First, it describes a classically chaotic system, which can be demonstrated through both analytical approaches and numerical methods. Intuitively, as shown in Fig. \ref{fig:Vxy}, \( V_{\ss R} \) possesses a double-well structure, which generally induces chaotic dynamics \cite{Holmes1979}. The two wells correspond to the two zeros of \( V_{\ss R} \), located at \( (x, y) = (0, 0) \) and \( (1, 0) \).

    Second, the denominator in \( V_{\ss R} \) exhibits a number-theoretic structure, since the inverse of \( \tau(z) \) is given by
    \begin{align}
        z(\tau) = \left( 1 - \frac{E_4^3(\tau)}{E_6^2(\tau)} \right)^{-1} \ , \label{equ:tau2z}
    \end{align}
    as provided by Ramanujan in his notebook \cite{Berndt1998}, where \( E_4(\tau) \) and \( E_6(\tau) \) are the Eisenstein series defined by
    \begin{align}
        E_{2k}(\tau) & := \frac{1}{2 \zeta(2k)} \sum_{\substack{(m, n) \in \mathbb{Z}^2 \\ (m, n) \neq (0, 0)}} \frac{1}{(m + n \tau)^{2k}} \ .
    \end{align}
    By the modularity of Eisenstein series, one obtains \cite{Diamond2005}
    \begin{align}
        z(\bv{\gamma} \tau) = z(\tau) \ , \label{equ:invariant}
    \end{align}
    where \( \bv{\gamma} = \left( \begin{array}{cc} a & b \\ c & d \end{array} \right) \in SL(2, \mathbb{Z}) \) and we define the action of \( \bv{\gamma} \) on \( z \) by \( \bv{\gamma} z := \frac{a z + b}{c z + d} \). Note that this property of \( z(\tau) \) is key to our construction.

    Finally, the numerator of \( V_{\ss R} \) is designed to encode the functional symmetry of the zeta function. This feature, together with the second point mentioned above, will become evident when we solve the eigenenergy problem in the next section.

	\textit{The eigenenergies}-
	The time-independent Schr\"odinger equation corresponding to the geometric potential (\ref{equ:VP}) is
    \begin{align}
        \hat{H}_{\ss R} \psi = E \psi \ \label{equ:seq}, 
    \end{align}
    where \( \hat{H}_{\ss R} := - V_{\ss R}^{\frac{1}{2}} \hat{\Delta} V_{\ss R}^{\frac{1}{2}} \) is the Hamiltonian. Note that, in general, we assume the eigenenergy \( E \in \mathbb{C} \). Furthermore, to avoid singularities near the zeros of \( V_{\ss R} \), we impose the requirement that the wave function is everywhere bounded on \( \mathbb{R}^2 \); that is, there exists a constant \( M \in \mathbb{R}^+ \) such that
    \begin{align}
        |\psi(x, y)| < M \label{equ:bc}
    \end{align}
    holds for all \( (x, y) \in \mathbb{R}^2 \). We will now prove that the problem defined by Eq.~(\ref{equ:seq}) and (\ref{equ:bc}) admits scattering state solutions, where the eigenenergy is discrete, and the energy levels satisfy \( E_n = \rho_n (1 - \rho_n) \).

	\begin{figure*}
		\centering
		\subfigure[]
		{
			\includegraphics[width= 0.45 \linewidth]{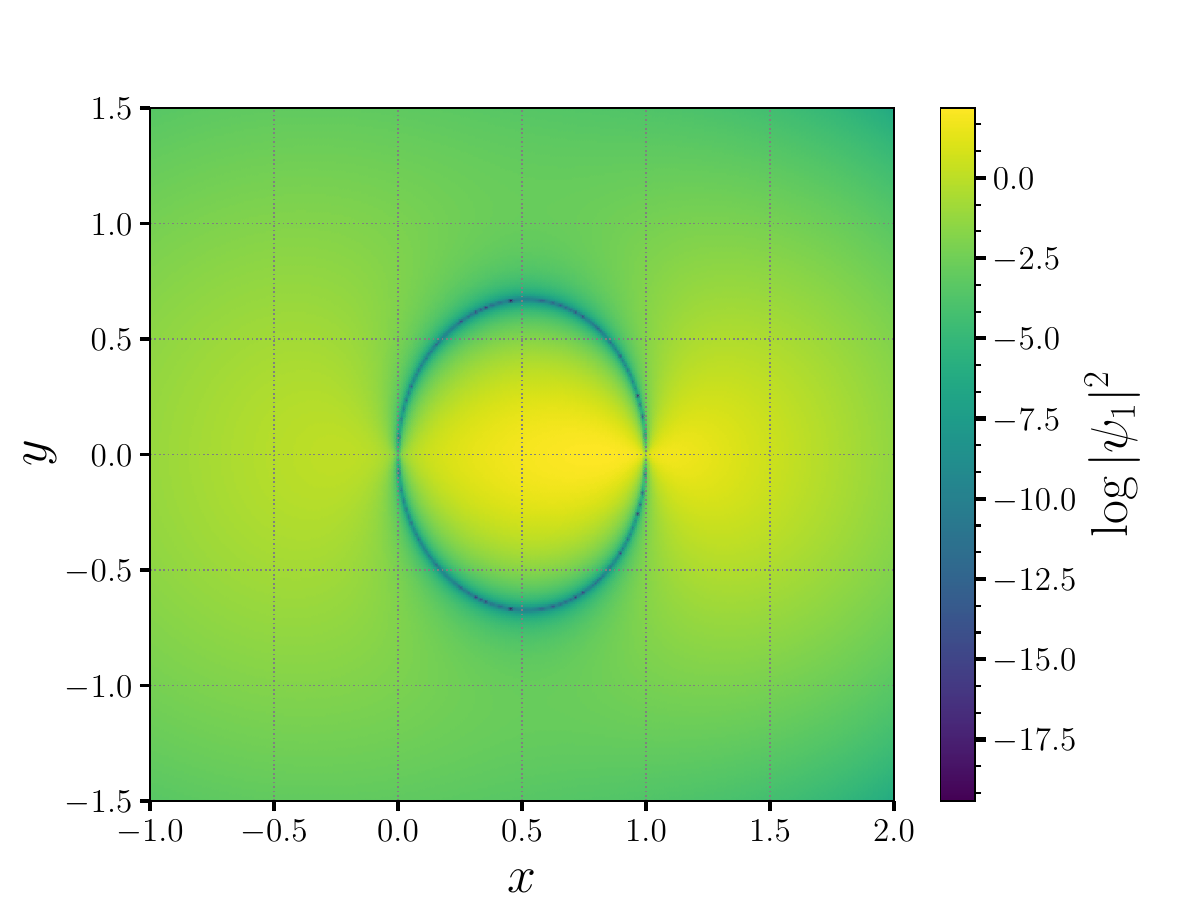}
			\label{fig:psi1}
		}
		\subfigure[]
		{
			\includegraphics[width= 0.45 \linewidth]{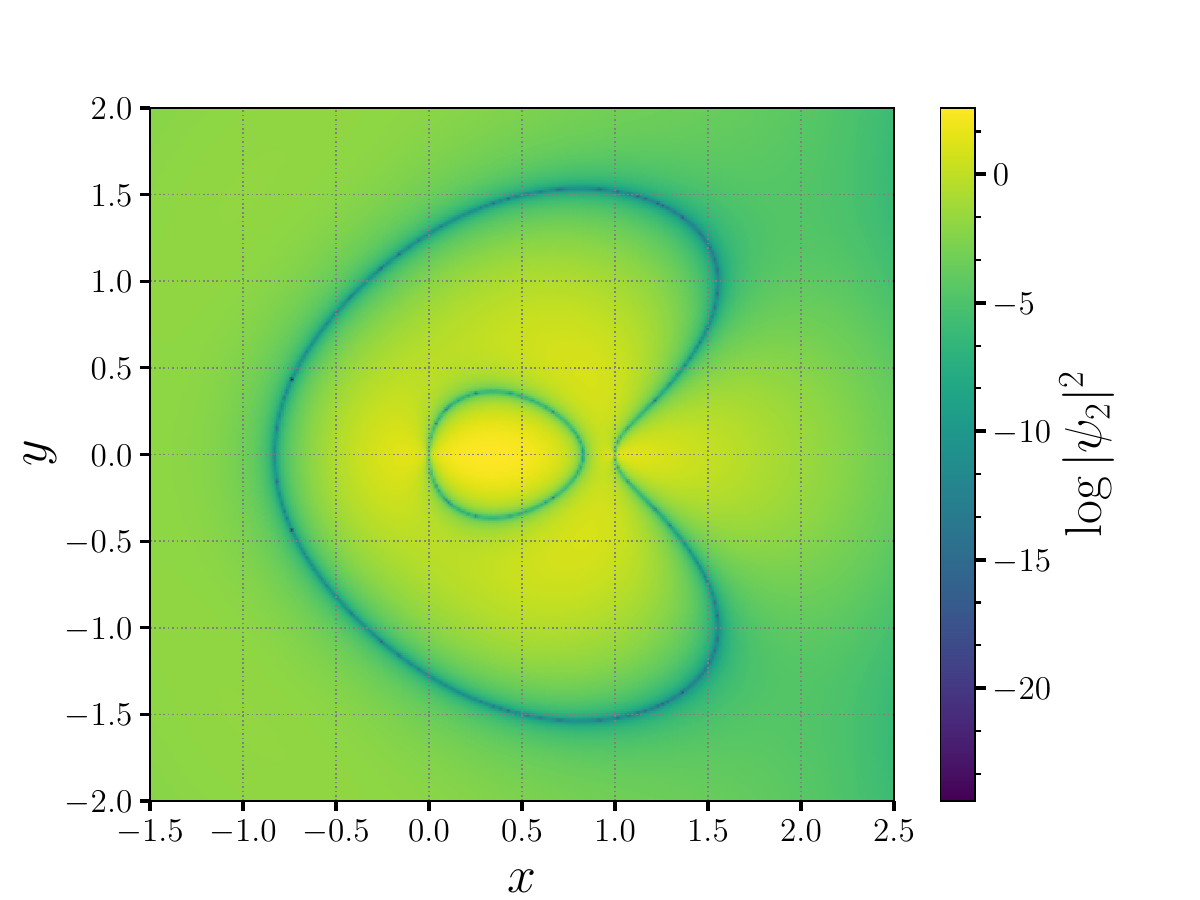}
			\label{fig:psi2}
		}
		\caption{The probability density of the first two eigenstates, corresponding to the first two Riemann zeros: (a) \( \rho_1 = 0.5 + \im 14.13472\ldots \) and (b) \( \rho_2 = 0.5 + \im 21.02203\ldots \), respectively, is shown. At the zeros of \( V_{\ss R} \), namely \( z = 0 \) and \( z = 1 \) (where \( z := x + \im y \)), the wave function indeed vanishes. We plot the logarithm of the probability density so that the nodal line, on which the wave function \( \psi(z) \) vanishes, is clearly visible.
        } \label{fig:psi} 
	\end{figure*}

    Using complex notation and changing the variables from \( z \) to \( \tau \), the Schr\"odinger equation (\ref{equ:seq}) can be written in a much simpler form:
	\begin{align}
		(\tau - \tau^* ) ^2 \frac{\partial} { \partial \tau}  \frac{\partial}{ \partial \tau^* }  \phi(\tau)  = E  \phi(\tau) \ ,   \label{equ:seq2}
	\end{align}
    where $^*$ denotes the complex conjugate and we have defined the \textit{reduced wave function} $\phi(\tau) := \psi(z(\tau))V^{\frac{1}{2}}(z(\tau)) $. 
	
    However, the coordinate transformation  \(z \to \tau\) is not one-to-one;  multiple values of \(\tau\)  correspond to the same $z$.
	Therefore, additional constraints must be imposed to ensure that the values of $\phi$ are identical at all $\tau$s corresponding to the same $z$.  
	Specifically, by Eq.~(\ref{equ:invariant}) and the definition of $\phi(\tau) $ one has 
	\begin{align}
		\phi(\tau) = \phi(\bv{\gamma} \tau)\ .  \label{equ:mc}
	\end{align}
    
    Furthermore, we must analyze the wave function \( \psi(z) \) near the zeros of \( V_{\ss R} \), specifically at \( z = 0 \) and \( z = 1 \), to check for potential singularities. An asymptotic analysis near $z=0$ shows that (see Section III of Supplemental Material\cite{SM})
    \begin{align}
        \psi(z) = \left[\phi(\im ) + O(|z|)\right]\left[ C |z|^{-\frac{1}{2}} + O(1) \right]\ ,\label{equ:app:i}
    \end{align}
    where $C$ is a non-zero constant. To satisfy boundary condition ($\ref{equ:bc}$), it is both sufficient and necessary that $\phi(\im) = 0 $. A similar asymptotic analysis for $z = 1 $ gives $\phi(e^{\im \frac{\pi}{3} }) = 0$.
    In conclusion, we have the boundary conditions
	\begin{align}
		\phi(\im ) = \phi(e^{\im \frac{\pi}{3}}) = 0 \label{equ:bc2}
	\end{align}
    Note that $\im $ and $e^{\im \frac{\pi}{3}}$ occur here because the coordinate transformation $z \to \tau$ maps the zeros of \(V_{\ss R}\) $z = 0 $ and $z =1 $ to $\tau = \im $ and $\tau = e^{\im \frac{\pi}{3}} $ respectively.
	
	Now, our task is to solve the problem defined by Eq.~\eqref{equ:seq2}, Eq.~\eqref{equ:mc} and Eq.~ \eqref{equ:bc2}. We first find \( \phi^{(0)}_s(\tau) = (\Im \tau)^s  \) as the solution to Eq.~(\ref{equ:seq2}), although it fails to satisfy the modular condition \eqref{equ:mc}. To resolve this, we modify \( \phi_s^{(0)} \) by shifting and summing over the appropriate group elements as follows:
    \begin{align}
        \phi_s(\tau) = \sum_{\bv{\gamma} \in \Gamma_{\infty} \backslash SL(2, \mathbb{Z})} \phi^{(0)}_s(\bv{\gamma} \tau) \ , \label{equ:phi_s}
    \end{align}
    where
    \(
    \Gamma_{\infty} := \left( \begin{array}{cc} 1 & \mathbb{Z} \\ 0 & 1 \end{array} \right)
    \)
    is a subgroup of \( SL(2, \mathbb{Z}) \), and \( \Gamma_{\infty} \backslash SL(2, \mathbb{Z}) \) is the left quotient group.

	Through direct calculation, it can be shown that \( \phi_s \) satisfies the modular condition in Eq.~\eqref{equ:mc} and serves as an eigenstate of Eq.~\eqref{equ:seq} with eigenenergy \( E = s(1 - s) \). Moreover, it has a zeta function representation (see Section II of Supplemental Material \cite{SM}):
    \begin{align}
        \phi_s(\tau) = \phi_s^{(0)}(\tau) \frac{\zeta_{\ss E}(s, \tau)}{\zeta(2s)} \ , \label{equ:phi_t}
    \end{align}
    where the Epstein zeta function is defined by
    \begin{align}
        \zeta_{\ss E}(s, \tau) = \sum_{\substack{(m, n) \in \mathbb{Z}^2 \\ (m, n) \neq (0, 0)}} \frac{1}{|m \tau + n|^{2s}} \ . \label{equ:EZ}
    \end{align}

	Note that although the summation in Eq.~(\ref{equ:EZ}) converges only for \( \Re s > 1 \), it can be analytically continued to the entire complex plane \( \mathbb{C} \), except for a pole at \( s = 1 \). Additionally, since \( \zeta(s) \) has a pole at \( s = 1 \), we require \( s \neq \frac{1}{2} \), otherwise \( \phi_s(\tau) \) would degenerate into the zero solution.

	In the final step, we require the solution \( \phi_s(\tau) \) to satisfy Eq.~(\ref{equ:bc2}). We have
    \begin{align}
        \frac{\zeta_{\ss E}(s, \im)}{\zeta(2s)}  = \frac{\zeta_{\ss E}(s, e^{\im \frac{\pi}{3}})}{\zeta(2s)}  = 0 \ . \label{equ:zetaE}
    \end{align}
    By the analytic formulation of Gauss's quadratic reciprocity law, the two terms \( \zeta_{\ss E}(s, \im) \) and \( \zeta_{\ss E}(s, e^{\im \frac{\pi} {3}}) \) can be decomposed as the product of the Riemann zeta function and Dirichlet \( L \)-functions \cite{Zagier1981}:
    \begin{align}
        \zeta_{\ss E}(s, \im) &= 4 \zeta(s) L(\chi_{-4}, s) \nonumber \\ 
        \zeta_{\ss E}(s, e^{\im \frac{\pi}{3}}) &= 6 \zeta(s) L(\chi_{-3}, s) \ , \label{equ:dec}
    \end{align}
    where \( \chi_{-4} \) and \( \chi_{-3} \) are Dirichlet characters defined by
    \begin{align}
       \chi_{-m}(n) := 
        \left\{
        \begin{array}{cc}
            1 & n \equiv 1 \mod m \\
            -1 & n \equiv m-1 \mod m  \\ 
            0 & \text{otherwise}
        \end{array}
        \right. \nonumber  .
    \end{align}

    Since both characters are odd (i.e., \( \chi_{-4}(-1) = \chi_{-3}(-1) = -1 \)), the trivial zeros of \( L(\chi_{-4}, s) \) and \( L(\chi_{-3}, s) \) are located at negative odd integers, while the trivial zeros of \( \zeta(s) \) are located at negative even integers. Surprisingly, these trivial zeros are completely canceled out by the trivial zeros of \( \zeta(2s) \) at negative integers.
    Thus, Eq.~(\ref{equ:zetaE}) and (\ref{equ:dec}) imply that \footnote{Here, we assume that \( L(\chi_{-4}, s) \) and \( L(\chi_{-3}, s) \) do not share any non-trivial zeros. If they do, the common zeros would correspond to additional solutions in our model; however, these would not affect our main discussion.}
    \begin{align}
        \zeta(s) = 0, \quad 0 < \Re s < 1 \ .
    \end{align}
    Namely, by taking \( s \) to be the non-trivial zeros of the Riemann zeta function, the boundary condition (\ref{equ:bc}) can be satisfied. As a consequence, the eigenenergy is discrete, and the energy levels are given by \( E_n = (1 - \rho_n) \rho_n \).

	The asymptotic behavior of the obtained wave function \( \psi_n \) at infinity is crucial, as it determines the space in which the wave function resides. 
	As \( |z| \to \infty \), we have (see Section III of Supplemental Material\cite{SM})
    \begin{align}
        \psi_n(z) &= A_n \frac{1}{|z|} \log(|12^3 z|)^{-\frac{1}{2} + d_n} \Omega_{d_n}(|z|) \nonumber \\
        &+ o\left(\frac{1}{|z|} \log(|z|)^{-\frac{1}{2} + d_n}\right) \ , \label{equ:asy}
    \end{align}
    where \( \Omega_{d_n}(|z|) \) is a bounded oscillatory term given by
    \begin{align}
        \Omega_{d_n}(|z|) := \left\{
        \begin{array}{cc}
            \log(12^3 |z|)^{\pm \im \omega_n} & \text{if } d_n > 0 \\
            \cos(\omega_n \log \left(\log(12^3 |z|)/(2\pi)\right) - \varphi_n) & \text{if } d_n = 0
        \end{array}
        \right. \ . \nonumber
    \end{align}
    Here, \( A_n \neq 0 \) and \( \varphi_n \in \mathbb{R} \) are constants depending on \( \rho_n \), \( d_n := |\Re \rho_n - 1/2| \) is the distance of the zeta zeros \( \rho_n \) to the line \( \Re s = 1/2 \), and \( \omega_n = \Im \rho_n \) is the imaginary part of \( \rho_n \).

	As expected, the wave functions decay to zero as \( |z| \to \infty \). The decay rate depends on the distance \( d_n \), while the imaginary part of \( \rho_n \) (i.e., \( \omega_n \)) acts as the frequency. However, the decay rate is not fast enough for \( \psi_n \) to be a bound state. One can verify that \( \int_{\mathbb{R}^2} |\psi_n|^2 \, dx \, dy \to \infty \), which implies that the wave function \( \psi_n \) represents a scattering state.

    To gain an intuitive understanding of the wave function, we numerically calculate the first two eigenstates and present them in Fig.~\ref{fig:psi}. As expected, the two wave functions vanish at the zeros of \( V_{\ss R} \), namely \( (0, 0) \) and \( (1, 0) \), which is also consistent with our expectations(see Eq.~\eqref{equ:app:i}).

	\textit{Discussion}-  
	It can be seen that the reality of \( E_n \) directly implies the RH. 
    Writing
    \(
    \rho_n = \tfrac{1}{2} + \im \gamma_n ,
    \)
    we obtain
    \(
    E_n = \tfrac{1}{4} + \gamma_n^2 ,
    \)
    where \(\gamma_n \in \mathbb{C}\) but cannot be purely imaginary. 
    Hence, the reality of \( E_n \) necessitates the reality of \(\gamma_n\), which is precisely equivalent to the RH.

    Mathematically, however, standard spectral theory guarantees that bound states (which reside in a Hilbert space \( \mathcal{H} \)) possess real eigenvalues. In contrast, the wave function \( \psi_n \) describes a scattering state in the space of bounded functions \( \mathcal{B}(\mathbb{R}^2) \), which falls outside the scope of traditional Hilbert space spectral theorems.

    Here, we provide a possible path to prove that \( E_n \) is real. 
   From the Schr\"odinger equation, one can derive the probability conservation equation (see Section IV of Supplemental Material\cite{SM}):
   \begin{align}
        \dot{Q}(t) = - \oint_{\partial U} \bv{J} \cdot \bv{n} \, dl \ , \label{equ:Q1}
    \end{align}
    where \( Q(t) \) is the probability of finding the particle in a region \( U \),  
    and \( \bv{J} \) is the probability current density vector.
    For a stationary wave function \( \psi(z) e^{-\im E t} \), the time derivative of \( Q(t) \) is given by
    \begin{align}
        \dot{Q}(t) = 2 \Im (E) e^{2 \Im(E) t} \int_U d^2z \, |\psi(z)|^2 \ , \label{equ:Q2}
    \end{align}
    and the the probability current density vector is given by 
    \begin{align}
        \bv{J} := e^{2 \Im(E) t} \left(  \psi^* \hat{\bv{v}} \psi - \psi \hat{\bv{v}} \psi^* \ \right) , \label{equ:J}
    \end{align}
    with the velocity operator defined by \( \hat{\bv{v}} := -\im V_{\ss R}^{\frac{1}{2}} \nabla V_{\ss R}^{\frac{1}{2}} \).

    For each \( \psi_n \), suppose there exists a compact region \( U_n \subset \mathbb{C} \) (which depends on \( n \)) such that  
    1. \( \psi_n |_{\partial U_n} \equiv 0 \);  
    2. \( \int_{U_n} d^2 z \, |\psi_n|^2 \neq 0 \).  
    Then, by Eq.~\eqref{equ:Q1}--\eqref{equ:J}, it follows that \( \Im E_n = 0 \). Since the wave function \( \psi_n \) is real analytic, it cannot vanish almost everywhere on any compact region, and thus the second condition is automatically satisfied. Consequently, to establish the reality of the eigenenergy, it suffices to identify a closed nodal line of the corresponding wave function, where a nodal line is defined as a curve along which the wave function vanishes.

    We computed a number of wave functions $\psi_n$, and for each of them, a closed nodal line can be identified. For example, the "egg-shaped" curve shown in Fig.~\ref{fig:psi1} and the "heart-shaped" curve shown in Fig.~\ref{fig:psi2} serve as illustrations.

    We now turn to the theoretical possibility of proving the existence of such nodal lines.
    
    According to Eq.~\eqref{equ:bc2}, \( \tau = \im \) and \( \tau = e^{\im \frac{\pi}{3}} \) are two zeros of the reduced wave function \( \phi_{\rho_n}(\tau) \). Furthermore, one can prove that near \( \tau = \im \), the reduced wave function \( \phi_{\rho_n} \) can be expressed as (see Section III of Supplemental Material\cite{SM}):
    \begin{align}
        \phi_{\rho_n}(\im + \eta) &= \alpha_n (\eta^2 + {\eta^*}^2) + O(|\eta|^3)  \nonumber \\
        & = 2 \alpha_n \left( (\Re\eta)^2 - (\Im\eta)^2 \right) + O(|\eta|^3) \ , \label{equ:expp_in_i}
    \end{align}
    where \( \alpha_n \neq 0 \) is a constant depending on \( \rho_n \). If we ignore the higher-order terms, the nodal lines of the leading-order term \( (\Re\eta)^2 - (\Im\eta)^2 \) form a cross with \( \tau = \im \) as the vertex.
    Although this cross may be disrupted by higher-order terms (for example, \( \phi(\im + \eta) = (\Re \eta)^2 - (\Im \eta)^2 + \im (\Re \eta)^3 \)), we can assume its existence and then explore what conclusions can be drawn. 
    
    When we map the \( \tau \)-plane to the \( z \)-plane via Eq.~\eqref{equ:tau2z}, the four edges of this cross merge into two edges with \( z = 0 \) as the vertex. For the other point, \( \tau = e^{\im \frac{\pi}{3}} \), the situation is entirely analogous; thus, we may also assume two nodal lines with \( z = 1 \) as the vertex.
    
    Any nodal line that originates at \( z = 0 \) or \( z = 1 \) can only terminate at one of the points \( z = 0 \), \( z = 1 \), or \( z = \infty \). By the asymptotic behavior given in Eq.~\eqref{equ:asy}, such nodal lines cannot end at \( z = \infty \). Hence, only two scenarios remain possible:  1. Two nodal lines originate at \( z = 0 \) and terminate at \( z = 1 \);  2. One nodal line originates and terminates at \( z = 0 \), while another originates and terminates at \( z = 1 \).  
    Examples of the first and second scenarios can be illustrated in Fig.~\ref{fig:psi1} and Fig.~\ref{fig:psi2}, respectively. 
    In both cases, the existence of at least one closed nodal line is guaranteed. 
    We note that a nodal line may intersect itself or other nodal lines; however, this does not affect our conclusion.


   The remaining issue is that one needs to conduct a detailed analysis of all the higher-order terms in Eq.~\eqref{equ:expp_in_i} to ensure that they do not disrupt the cross. We numerically calculate a few wave functions and find that the cross exists in each of them(see Section IV of Supplemental Material \cite{SM}).  However, significant effort is still required to analytically confirm the existence of the cross for all wave functions, and we leave this for future work.

	In conclusion, we have constructed a novel Hamiltonian. Under the requirement that the wave function is bounded, this Hamiltonian admits \( E_n = \rho_n(1 - \rho_n) \) as the eigenenergies of scattering states. Our construction offers a fresh physical perspective on both the RH and HPC. It suggests that the non-trivial zeros of the Riemann zeta function may be more closely associated with scattering states rather than bound states. Moreover, the study of scattering states---particularly bounded scattering states---could serve as a potential avenue for proving the RH.

	\begin{acknowledgments}
		The author would like to thank Micheal V. Berry, Xi Kang and Tingfei Li  for useful discussion. We thank the anonymous reviewers for their valuable comments and suggestions, which have not only helped enhance the rigor of the paper but also increased the depth of its research.  We acknowledge the support from the National Key Research and Development Program of China (No. 2022YFA1602903, No. 2023YFB3002502), NSFC (No. 12533007); National Natural Science Foundation of China Regional Project under Grant 62262074; Science and Technology Plan Project of of Yunnan Province under Grant 202405AC350083.
	\end{acknowledgments}
	 
	\nocite{*}
	
	\bibliography{apssamp}

	\clearpage
	\includepdf[pages={{}, -}]{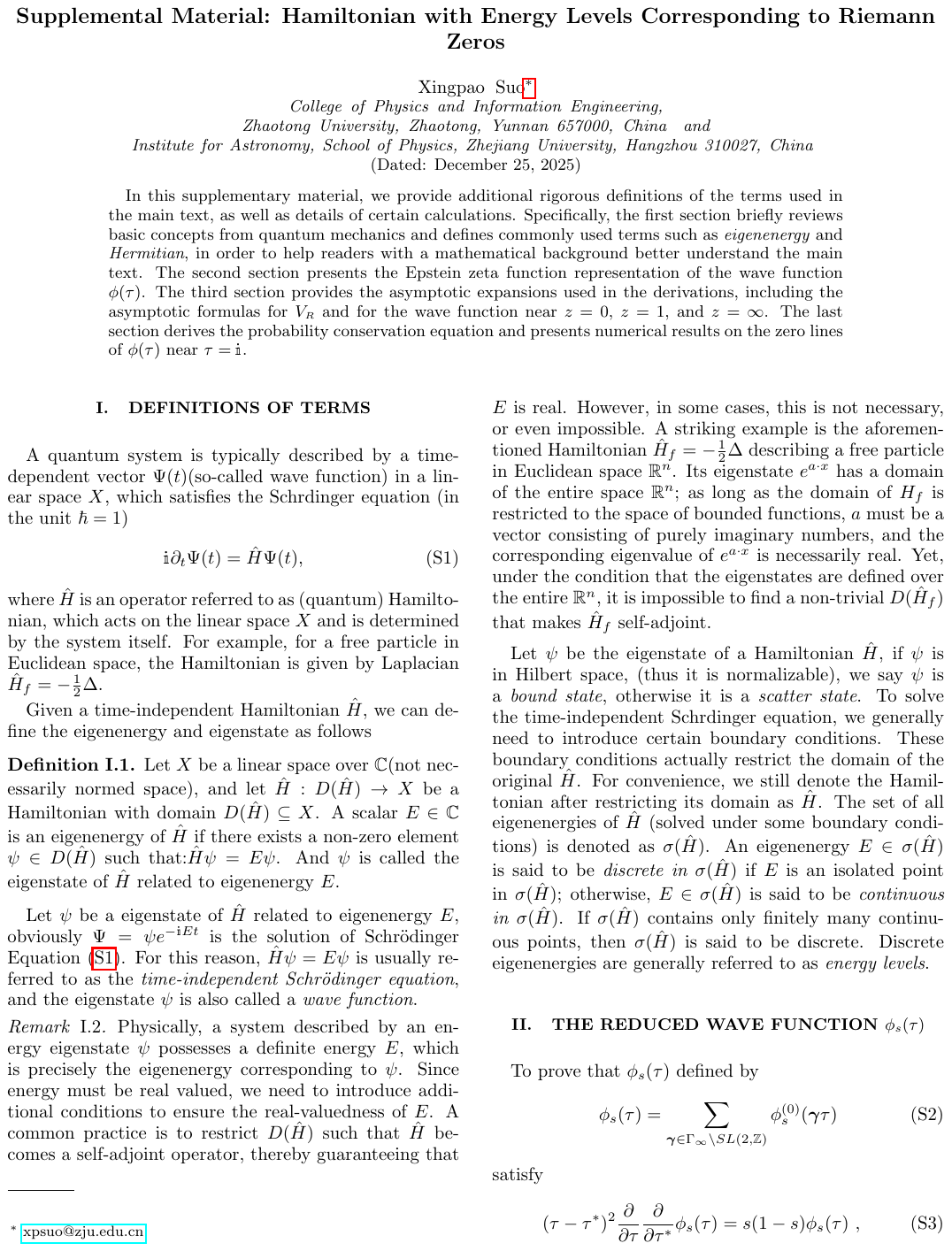}
	\pagestyle{empty}
	\clearpage
	
\end{document}